\documentclass[USenglish,twocolumn]{article}

\ifx\directlua\undefined\ifx\XeTeXcharclass\undefined
  \usepackage[utf8]{inputenc}                           
  \else\RequirePackage[no-math]{fontspec}[2017/03/31]\fi 
  \else\RequirePackage[no-math]{fontspec}[2017/03/31]\fi 
\usepackage[sort&compress,square,numbers]{natbib}
\usepackage[big,online]{dgruyter}
\usepackage{changes}

\theoremstyle{dgthm}

\theoremstyle{dgdef}

\begin{document}

\articletype{Research Article}

\author[1]{Changming Huang}
\author*[2]{Ce Shang}
\author[3]{Yaroslav V. Kartashov} 
\author*[4]{Fangwei Ye}
\affil[1]{Department of Physics, Changzhi University, Changzhi, Shanxi 046011, China. \url{https://orcid.org/0000-0002-3520-4922}} %
\affil[2]{King Abdullah University of Science and Technology (KAUST), Physical Science and Engineering Division (PSE), Thuwal 23955-6900, Saudi Arabia. E-mail: shang.ce@kaust.edu.sa. \url{https://orcid.org/0000-0003-2207-7036}}
\affil[3]{Institute of Spectroscopy, Russian Academy of Sciences, 108840, Troitsk, Moscow, Russia. \url{https://orcid.org/0000-0001-8692-982X}}
\affil[4]{School of Physics and Astronomy, Shanghai Jiao Tong University, Shanghai 200240, China. E-mail: fangweiye@sjtu.edu.cn. \url{https://orcid.org/0000-0003-2263-9000}}
\title{Vortex solitons in topological disclination lattices}
\runningtitle{Vortex solitons in topological disclination lattices}
\abstract{The existence  of thresholdless vortex solitons trapped at the core of disclination lattices that realize higher-order topological insulators is reported. The study demonstrates the interplay between nonlinearity and higher-order topology in these systems, as the vortex state in the disclination lattice bifurcates from its linear topological counterpart, while the position of its propagation constant within the bandgap and localization can be controlled by its power. It is shown that vortex solitons are characterized by strong field confinement at the disclination core due to their topological nature, leading to enhanced stability. Simultaneously, the global discrete rotational symmetry of the disclination lattice imposes restrictions on the maximal possible topological charge of such vortex solitons. The results illustrate the strong stabilizing action that topologically nontrivial structures may exert on excited soliton states, opening new prospects for soliton-related applications.}
\keywords{Vortex solitons; Topological disclination lattices; Stability; Propagation dynamics}
\journalname{Nanophotonics}
\journalyear{2023}
\journalvolume{XXX}

\maketitle

\section{Introduction} 
Vortex solitons are localized, self-trapped states with nonzero orbital angular momentum. They were encountered in various physical systems such as nonlinear optical materials, Bose-Einstein condensates, polariton condensates, and plasmas~\cite{desyatnikov2005optical, kartashov2019review, malomed2019review, mihalache2021review, pryamikov2021review, kevrekidis2008review, fetter2010review, kevrekidis2016book, carusotto2013review}. Since phase singularity in vortex is a topologically stable object persisting even in the presence of perturbations, such states have important applications in tweezers \cite{padgett2011tweezers}, vortex microlasers \cite{zhang2020tunable}, and information encoding \cite{torner2005imaging, torner2007imaging}. Vortex solitons are ideal for optical logic gates in all-optical computing and communication \cite{Kavokin2022}. At the same time, being higher-order excited nonlinear states, they are prone to various dynamical instabilities. Different approaches have been proposed to stabilize them (see reviews \cite{kartashov2019review, malomed2019review}) that include the utilization of competing or nonlocal nonlinearities, rapid parameter variations, spin-orbit coupling, and various optical potentials including periodic lattices \cite{malomed2001vort, yang2003vort, neshev2004vort, fleischer2004vort, law2009vort}. Remarkably, when such potentials possess discrete rotational symmetry, they impose restrictions on the available charges of supported vortex solitons \cite{ferrando2005vort, kartashov2005soliton, dong2022vortex}.
While it is predicted that semi-vortex solitons can emerge in the bulk of topological lattices in the continuum limit \cite{PhysRevA.98.013827, li2022topological} and non-vortical solitons sustained by continuous Jackiw-Rossi-like distortion \cite{nedic2023nonlinearity} were reported, strongly localized vortex solitons in potentials belonging to the class of topological insulators have not been studied to our knowledge.

The remarkable property of topological insulators is the existence of localized states at their edges or corners, which are protected by the system's topology. These states have energies that fall within the forbidden topological gaps. The theory of quantized polarization \cite{PhysRevLett.62.2747, PhysRevB.48.4442} connecting the topological properties of bulk bands in such structures with the appearance of edge states has recently been extended from dipole to multipole moments, showcasing the development from first-order topological insulators \cite{RevModPhys.82.3045, RevModPhys.83.1057} to higher-order ones \cite{Benalcazar61, PhysRevB.96.245115, Schindler2018np, Peterson2018, Xue2018, Mittal2019, PhysRevResearch.3.013239, Xie2021}. The bulk-boundary correspondence in these systems may be characterized by a co-dimension ranging from one to higher. Furthermore, topological systems exhibit a rich variety of nonlinear phenomena that acquire unique features due to their topological nature. These phenomena include the formation of topological solitons \cite{lumer2013sol, ablowitz2014sol, leykam2016edge, kartashov2016modulational, li2018lieb, mukherjee2020sol, maczewsky2020nonlinearity, xia2020sol, mukherjee2021sol, kirsch2021nonlinear, hu2021sol, fleury2019sol, smirnova2019sol, kartashov2022trim}, lasing in topological states \cite{bahari2017nonreciprocal, harari2018topological, bandres2018topological, kartashov2019two}, enhanced generation of higher harmonics \cite{kruk2021harmonics}, nonlinear Thouless pumps \cite{jurgensen2021quantized, fu2022nonlinear, jurgensen2022chern, fu2022two, jurgensen2023quantized}, among others. Despite the exciting opportunities provided by topological systems for the formation of fundamental topological solitons, the absence of discrete rotational symmetry at the boundaries of most of such topological structures poses a significant challenge for creating topological vortex solitons.

This work aims to introduce optical vortex states in a nonlinear version of recently discovered topological disclination lattices \cite{teo2013existence, benalcazar2014classification, PhysRevB.101.115115, peterson2021, liu2021disclination, wu2021disclination,hwang2023vortex} (also explored in acoustic realizations \cite{wang2020acoustic, chen2022acoustic, deng2022acoustic}). We find that these solitons form at the disclination core of the lattices with different discrete rotational symmetries, that they are thresholdless because they bifurcate from linear topological disclination modes, and that lattice topology may grant enhanced stability to such states. Our results not only provide the first example of compact nonlinear vortex state in a topological system, but they also show that such states in disclination lattices can possess topological charges forbidden in systems based on periodic lattices, such as square or honeycomb ones.

\section{Disclination lattices}
\begin{figure}[t]
\centering
\includegraphics[width=1\columnwidth]{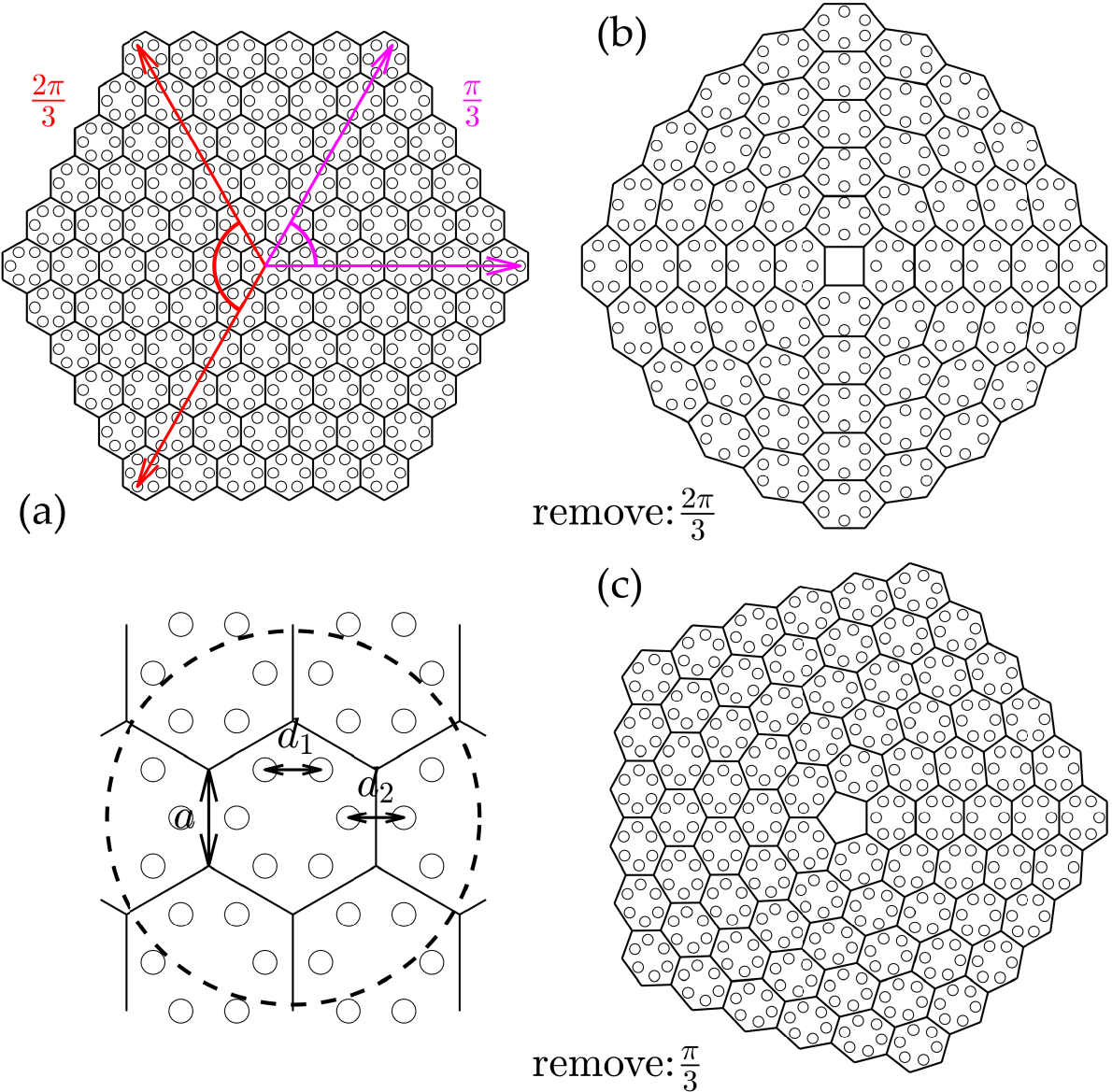}
\caption{Illustration of the method of construction of disclination lattice. (a) Original hexagonal lattice.
 (b),(c) Disclination lattices with $\mathcal{C}_4$ (or $\mathcal{C}_5$) symmetry obtained by removing a $2\pi/3$ (or $\pi/3$) sector from the original hexagonal
lattice.}
\label{fig1}
\end{figure}

To generate a disclination lattice, we employ the Volterra process, which involves removing or inserting a $n\pi/3$ section from a hexagonal sample \cite{teo2013existence}. This process generates a disclination with a Frank angle of $\pm n\pi/3$, resulting in structures with $\mathcal{C}_{6\pm n}$ discrete rotational symmetry, as shown in Fig.~\ref{fig1}. For the optical realization, we suppose that original $\mathcal{C}_6$ sample is composed of Gaussian waveguides with the depth $p$ and width $w$ [see the inset below Fig.~\ref{fig1}(a)], $a$ is the length of the lattice cell, $d_1$ is the intracell waveguide spacing, $d_2$ is the intercell waveguide spacing. We set $a \equiv 3$, $p=8$, and $w=0.5$. Varying parameter $\gamma=d_1/d_2$ quantifies relative strength of intra- and intercell coupling. As shown in Fig.~\ref{fig1}(b), one can obtain a lattice with  $\mathcal{C}_4$ discrete rotational symmetry by removing the $2\pi/3$ sector from the hexagonal structure and gluing the remaining parts together. Similarly, by inserting the $2\pi/3$ sector into a hexagonal structure, one can obtain a lattice with $\mathcal{C}_8$ symmetry (see Supplementary Materials \cite{suppmat}). Disclination lattices with $\mathcal{C}_5$ [Fig. \ref{fig1}(c)] and $\mathcal{C}_7$ rotational symmetry are obtained by removing or inserting the $\pi/3$ sector. In all cases, one can see the formation of a disclination core in the center of the lattice. Remarkably, this procedure allows to construct structures with discrete rotational symmetry not attainable in usual periodic lattices.

The propagation of light beam along the $z$-axis in the disclination lattice created in the cubic nonlinear medium can be described by the nonlinear Schr\"odinger equation for dimensionless field amplitude $\Psi$:
\begin{equation}\label{eq1}
i\frac{\partial \Psi}{\partial z}=-\frac{1}{2}\nabla^2\Psi-\mathcal{V}(x,y)\Psi-g|\Psi|^2\Psi,
\end{equation}
where $\nabla=(\partial/\partial x,\partial/\partial y)$, the transverse $x,y$ and longitudinal $z$ coordinates are normalized to the characteristic scale $r_0=10~\mu\textrm{m}$ and diffraction length $kr_0^2\approx 1.14 ~\textrm{mm}$ respectively, $k=2\pi n/\lambda$ is the wavenumber at $\lambda=800~\textrm{nm}$, $n\approx 1.45$ is the background refractive index, the dimensionless intensity $|\Psi|^2$ corresponds to $I=n|\Psi|^2/k^2r_0^2|n_2|$ (in fused silica $n_2\approx 2.7 \times 10^{-20}~\textrm{m}^2/\textrm{W}$), $g=+1~(g=-1)$ corresponds to focusing (defocusing) nonlinearity. The function $\mathcal{V}(x, y)=p \sum_{m, n} e^{-[\left(x-x_m\right)^2+\left(y-y_n\right)^2] / w^2}$ describes a disclination lattice, where lattice depth $p=k^2r_0^2\delta n/n$ is proportional to the refractive index contrast $\delta n$ ($p=8$ corresponds to $\delta n\sim 9\times10^{-4}$), $(x_m,y_n)$ are the coordinates of the waveguides in disclination structure, and $w=0.5$ (corresponding to $5~\mu \textrm{m}$) is the waveguide width. Such structures can be inscribed in nonlinear transparent dielectrics using the fs-laser writing technique \cite{maczewsky2020nonlinearity, mukherjee2020sol}. Straight waveguides in such lattices may exhibit low propagation losses not exceeding $0.1\,\textrm{dB/cm}$ at $\lambda=800\,\textrm{nm}$, enabling observation of solitons and rich nonlinear dynamics on typical sample lengths of $20~\textrm{cm}$.

\begin{figure}[!hpt]
\centering
\includegraphics[width=1\columnwidth]{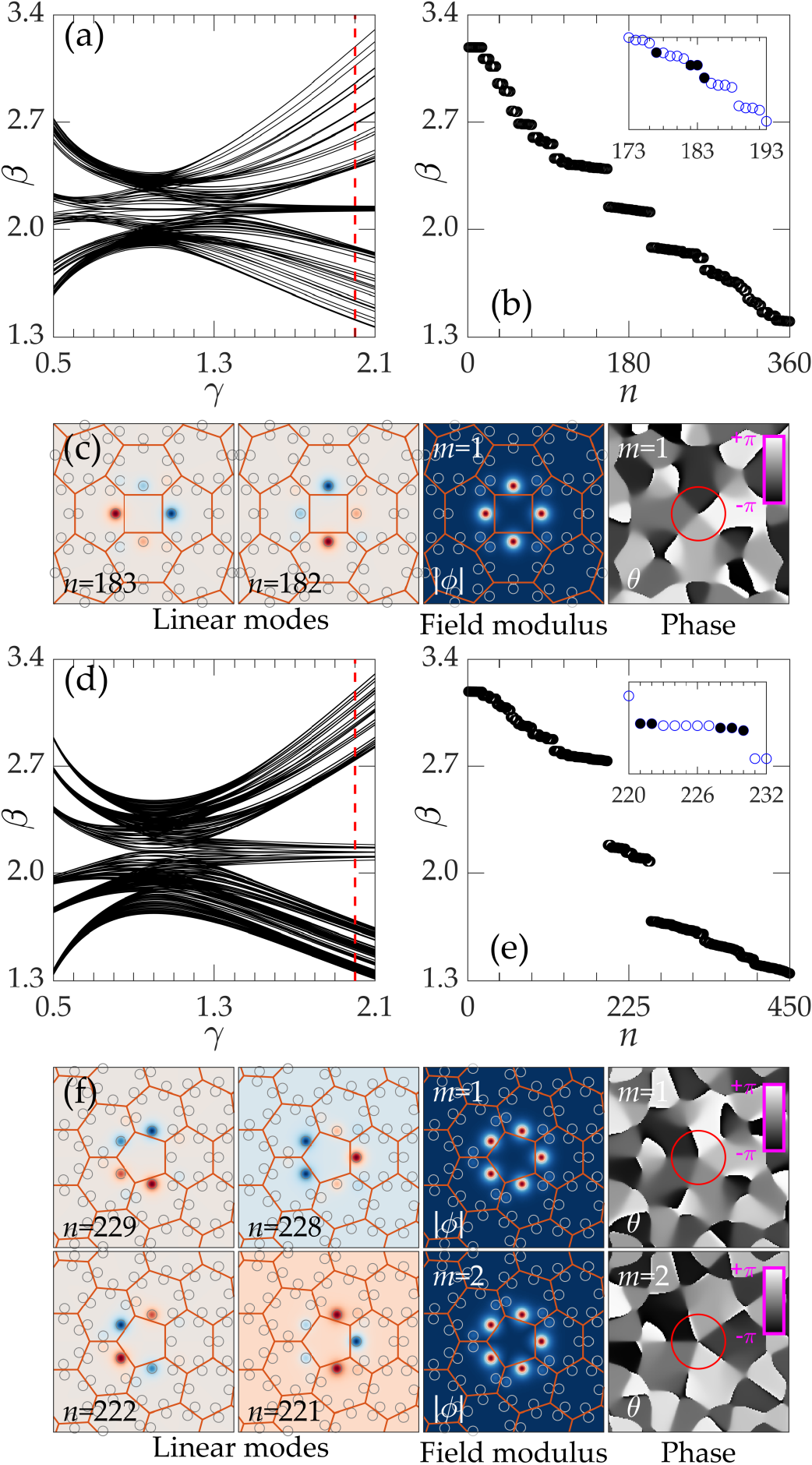}
\caption{(a) Dependencies $\beta(\gamma)$ for $\mathcal{C}_4$ lattice. (b) Eigenvalues at $\gamma=2$ corresponding to the red dashed line in (a). The inset in (b) shows the indices of four linear disclination modes localized at the disclination core. (c) Two localized degenerate modes supported by the $\mathcal{C}_4$ lattice and field modulus and phase distributions of vortex mode with $m=1$ that they generate. (d) Dependencies $\beta(\gamma)$ for $\mathcal{C}_5$ lattice. (e) Eigenvalues at $\gamma=2$ corresponding to the red dashed line in (d). Localized disclination modes are marked with black solid dots in the inset of (e). (f) Linear vortex mode with $m=1$ composed from the degenerate states $\phi_{n=228,229}$, and that with $m=2$ composed from the degenerate states $\phi_{n=221,222}$. Magenta lines in (c) and (f) depict lattice cells, while circles show waveguides.}
\label{fig2}
\end{figure}

\section{Linear disclination modes}
To understand the structure of possible vortex solitons in disclination lattices, we first examine their linear spectra, which can be obtained by setting $g=0$ in Eq.~(\ref{eq1}) and searching the eigenmodes of the form $\Psi(x,y,z)=\phi(x,y)e^{i\beta z}$, where $\beta$ is the propagation constant (eigenvalue) and the real function $\phi(x,y)$ describes the modal field. The dependencies of the eigenvalues $\beta$ on the distortion parameter $\gamma$ is shown for disclination lattices with $\mathcal{C}_4$ and $\mathcal{C}_5$ discrete rotational symmetry in Figs.~\ref{fig2}(a) and \ref{fig2}(d), respectively. In the non-trivial topological regime $\gamma>1$ (see \cite{suppmat} and  \cite{teo2013existence, benalcazar2014classification, PhysRevB.101.115115, peterson2021, liu2021disclination, wu2021disclination} for topological characterization of disclination lattices), localized modes emerge on the central disclination core. 
The spectral gap, where disclination modes appear, opens for sufficiently large $\gamma$ and increases with $\gamma$ leading to stronger localization of disclination states. For a fixed value of $\gamma=2$ [indicated by the red dashed lines in Figs.~\ref{fig2}(a) and~\ref{fig2}(d)], the gap of the $\mathcal{C}_4$ lattice corresponds to $\beta \in [1.884,2.397]$, while in $\mathcal{C}_5$ lattice it corresponds to $\beta \in [1.687,2.734]$. There are no localized modes in the non-topological regime $\gamma<1$.

The disclination lattice with $\mathcal{C}_N$ symmetry supports $N$ topological disclination modes, some of which can be degenerate [see insets of Figs.~\ref{fig2}(b) and~\ref{fig2}(e)]. Linear combination of degenerate modes can produce vortex disclination states. In the lattice with $\mathcal{C}_4$ symmetry only one set of degenerate states $\phi_{n=182,183}$ with identical eigenvalues was found (here $n$ is the index of disclination state depending on the structure size), whose linear combination $\phi_{n=182} \pm i\phi_{n=183}$ produces single-charge ($m=\pm1$) vortex mode, whose field modulus and phase distributions are shown in Fig. \ref{fig2}(c) (notice that this mode occupies all four sites of the disclination core). The topological charge of the mode is defined using the formula:
$m=\frac{1}{2\pi}\text{Im}\int_{0}^{2\pi} \partial\phi(r_0,\varphi)/\partial \varphi/\phi(r_0,\varphi)d\varphi$,
where $r_0$ is a fixed small radius, and $\varphi$ is the azimuthal angle.

In the spectrum of disclination lattice with $\mathcal{C}_5$ symmetry [see the example in Fig.~\ref{fig2}(e) for $\gamma=2$] one can identify two pairs of degenerate states $\phi_{n=228,229}$ and $\phi_{n=221,222}$. The combination $\phi_{n=229} \pm i\phi_{n=228}$ yields $m=\pm1$ disclination vortex, while combination $\phi_{n=222} \pm i\phi_{n=221}$ yields $m=\pm 2$ state, both of them are strongly localized on the disclination core for this value of $\gamma$, see profiles in Fig.~\ref{fig2}(f).  $\mathcal{C}_7$ and $\mathcal{C}_8$ structures with higher rotational symmetries (see \cite{suppmat}) support disclination vortices with charges up to $m=\pm 3$, and so on, \textcolor{black}{so that available charge of disclination vortex in $\mathcal{C}_N$ lattice is given by $m<N/2$ (for even $N$) and $m<(N+1)/2$ (for odd $N$).
It should be stressed that topological vortex modes on disclination are rather robust objects that persist even in the presence of disorder in the lattice. To illustrate this, we show that such vortices survive upon propagation in the lattice, where depths of individual waveguides were allowed to change randomly within the interval $[p-\delta,p+\delta]$, with $\delta\ll p$ (see \cite{suppmat}).}

\section{Vortex solitons in disclination lattices}
We now consider vortex solitons governed by Eq.~(\ref{eq1}) with $g\ne 0$. Because linear spectrum is characterized by the gap with topologically protected linear disclination modes in it, in-gap vortex solitons can bifurcate from such states in both focusing and defocusing media, and, importantly, nonlinearity can be used to control the location of such nonlinear states in the gap. To study properties of such "excited" vortex solitons, clearly different from their fundamental counterparts  \cite{ren2023nonlinear}, we search for solutions of Eq.~(\ref{eq1}) in the form $\Psi(x,y,z)=\phi(x,y)e^{i\beta z}$, where $\phi(x,y)=\phi_r(x,y)+i\phi_i(x,y)$ is the complex function describing vortex soliton profile, while $\beta$ is the nonlinear propagation constant. The real $\phi_r$ and imaginary $\phi_i$ parts satisfy coupled nonlinear equations
\begin{equation}\label{eq2}
-\frac{1}{2}\nabla^2 \phi_{r,i}-\mathcal{V}\phi_{r,i}-g(\phi_r^2+\phi_i^2)\phi_{r,i}+\beta \phi_{r,i}=0,
\end{equation}
that can be solved using a standard Newton iteration method with a targeted error tolerance $10^{-8}$.

\begin{figure}[t]
\centering
\includegraphics[width=1\columnwidth]{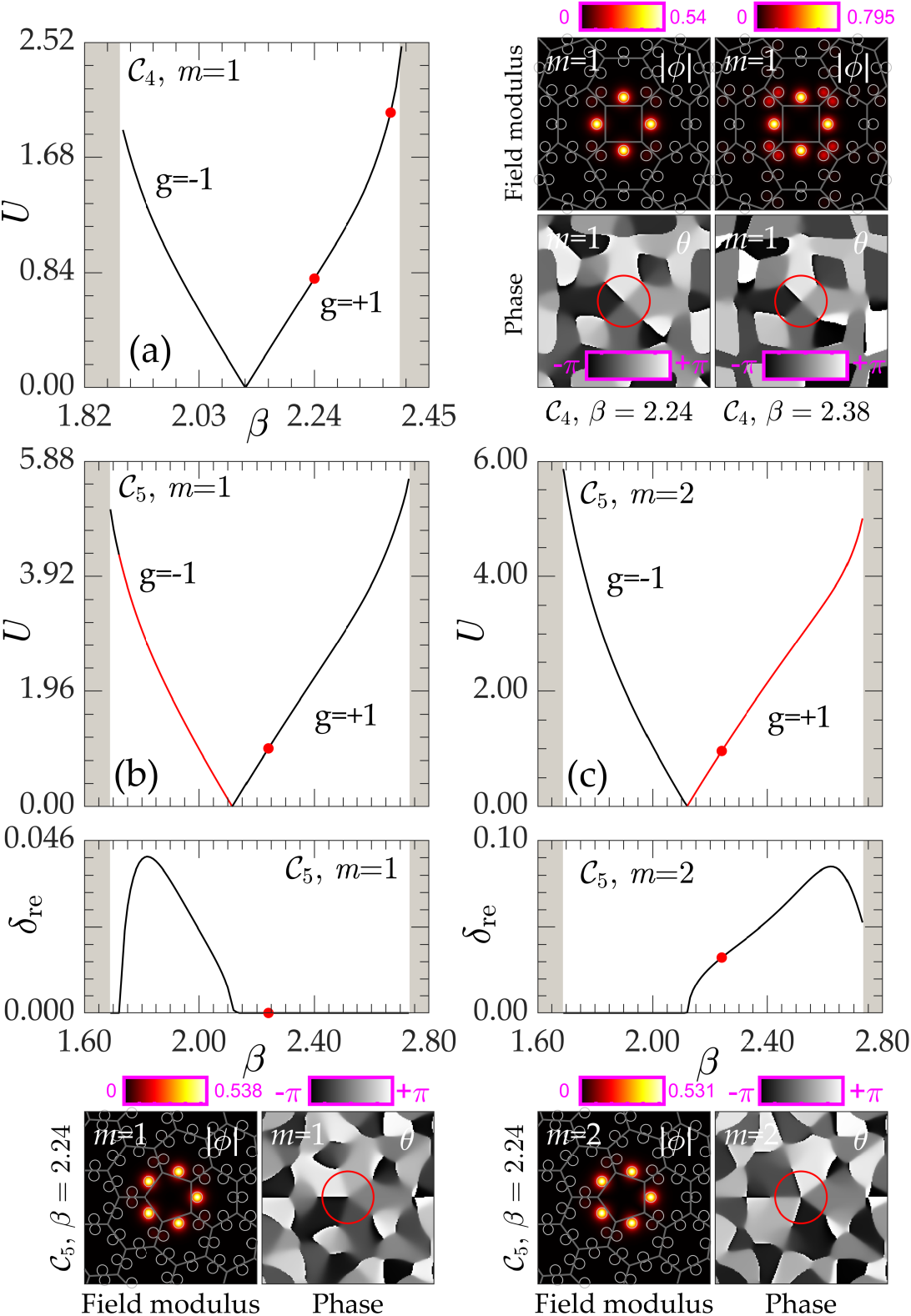}
\caption{Power $U$ and maximal (among all perturbations) real part of perturbation growth rate $\delta_\mathrm{re}$ versus propagation constant $\beta$ for vortex soliton families in disclination lattices with focusing and defocusing nonlinearity. (a) $m=1$, $\mathcal{C}_4$ lattice, (b) $m=1$, $\mathcal{C}_5$ lattice, and (c) $m=2$, $\mathcal{C}_5$ lattice. Field modulus and phase distributions of typical vortex solitons corresponding to the red dots are shown at the right (a) or at the bottom (b,c) of panels with $U(\beta)$ curves. The gray regions in the $U(\beta)$ plot illustrate bulk bands. Stable and unstable branches are indicated by black and red lines, respectively.}\label{fig3}
\end{figure}

\begin{figure*}[t]
\centering
\includegraphics[width=1.7\columnwidth]{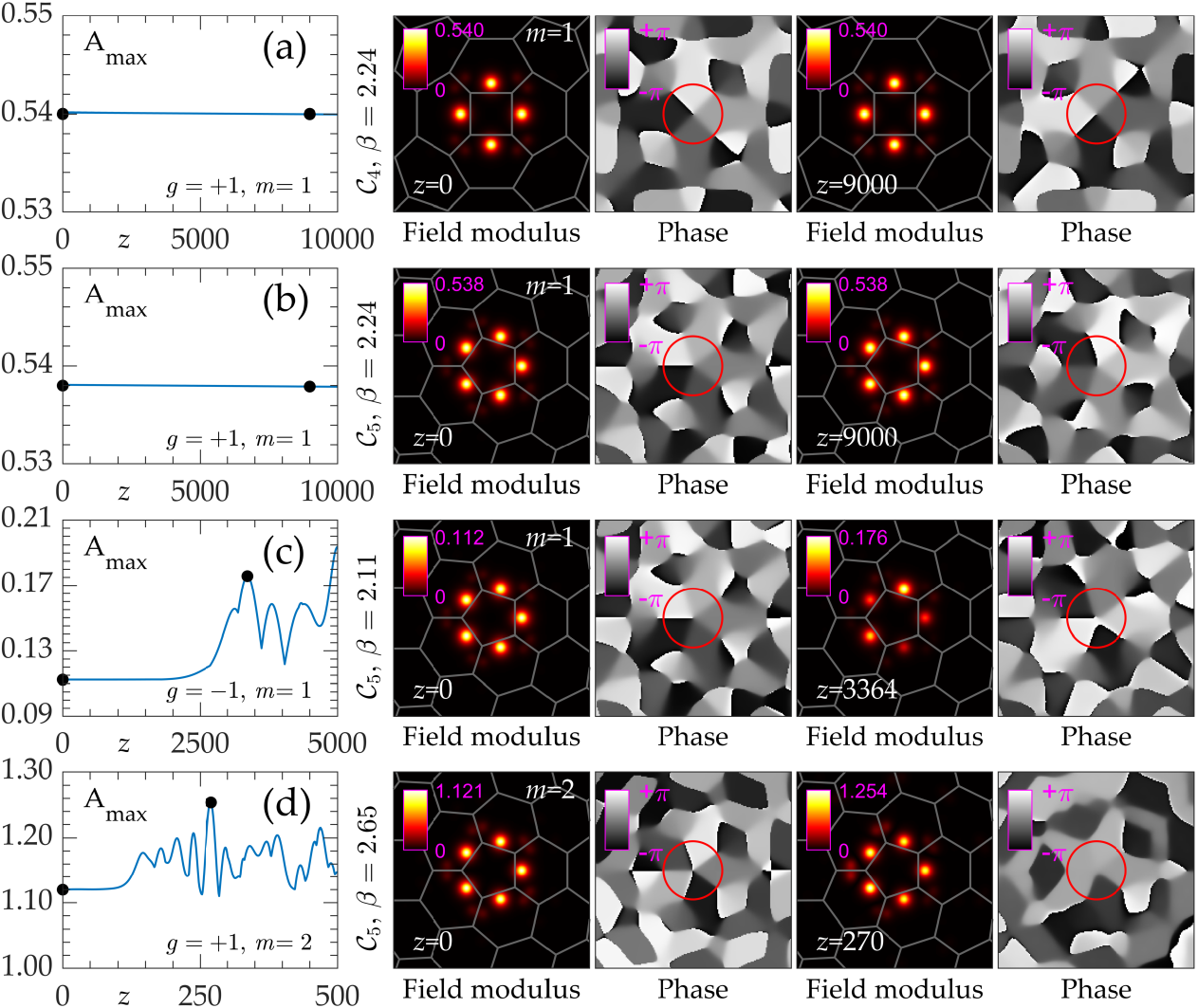}
\caption{Propagation dynamics of stable (a),(b) and unstable (c),(d) vortex solitons. Peak amplitude $A_\text{max}$ versus propagation distance $z$ is shown in the left plots, while field modulus and phase distributions corresponding to the black dots on $A_\text{max}(z)$ plots are displayed in the right panels. (a) $\mathcal{C}_4$ lattice, $\beta=2.24$, $m=1$, (b) $\mathcal{C}_5$ lattice, $\beta=2.24$, $m=1$, (c) $\mathcal{C}_5$ lattice, $\beta=2.11$, $m=1$, and (d) $\mathcal{C}_5$ lattice, $\beta=2.65$, $m=2$ in (d). In (a),(b),(d) $g=+1$, while in (c) $g=-1$.}
\label{fig4}
\end{figure*}

The families of vortex solitons in disclination lattices with $\mathcal{C}_4$ and $\mathcal{C}_5$ symmetry are presented in Fig.~\ref{fig3} in the form of dependencies of soliton power $U=\int_{-\infty}^{+\infty}\int_{-\infty}^{+\infty}|\phi(x,y)|^2dxdy$ on the propagation constant $\beta$.
The formation mechanism of vortex solitons in the topological gap is similar to that of gap solitons, with the only difference being that gap solitons bifurcate from the edge of the allowed band, while our disclination vortices bifurcate from the topological state within the depth of the gap.
For a given lattice $\mathcal{C}_N$ we present the results for focusing ($g=+1$) and defocusing ($g=-1$) medium in the same plot. It can be observed that the sign of nonlinearity determines the direction of bifurcation: in focusing medium soliton power $U$ increases with the increase in $\beta$, while in defocusing medium $U$ increases with a decrease in $\beta$. Power vanishes at the bifurcation point, which corresponds to the  $\beta$ value associated with degenerate linear modes producing vortex of a given charge $m$ (thus such solitons are thresholdless), and it notably grows as $\beta$ approaches upper or lower edges of the gap. While well within the gap vortex solitons are localized mostly on $N$ sites of disclination core, close to the gap edge they start expanding into the lattice bulk and, if $\beta$ shifts into the band, coupling with bulk modes occurs, and the soliton delocalizes. This is illustrated in Fig. \ref{fig3}(a) where profiles of two $m=1$ vortex solitons with different propagation constants are compared. While in $\mathcal{C}_4$ lattice only vortex solitons with $m=1$ were obtained, in $\mathcal{C}_5$ lattices we found $m=1$ [Fig. \ref{fig3}(b)] and $m=2$ [Fig. \ref{fig3}(c)] families (for properties of higher-charge nonlinear states in $\mathcal{C}_{7,8}$ lattices see~\cite{suppmat}). Field modulus and phase distributions for the latter states are presented below corresponding $U(\beta)$ distributions. Notice that in a nontopological lattice achieved by setting $\gamma<1$, no linear localized states are sustained within the gap, thus vortex solitons can only appear above some power threshold.

One of the most essential aspects for potential experimental realization is the stability of disclination vortex solitons and this is where they show properties strongly departing from properties of vortex solitons in nontopological lattices. A linear stability analysis and modeling of propagation were performed for the perturbed vortex solitons. We searched for perturbed solutions of Eq.(\ref{eq1}) in the form $\Psi(x,y,z)=[\phi_r(x,y)+i\phi_i(x,y)+u(x,y)e^{\delta z}+iv(x,y)e^{\delta z}]e^{i\beta z}$, where $u$, $v$ are real and imaginary parts of perturbation, respectively. Linearization of Eq.(\ref{eq1}) around $\phi_r$ and $\phi_i$ yields the eigenvalue problem:
\begin{equation}
\begin{array}{l}
\delta u = [-\frac{1}{2}{\nabla ^2}+\beta- g(\phi _r^2 + 3\phi _i^2)- {\cal V}]v- 2g{\phi _r}{\phi _i}u,\\
\delta v = [+\frac{1}{2}{\nabla ^2}-\beta+g(3\phi _r^2 + \phi _i^2)+{\cal V}]u  + 2g{\phi _r}{\phi _i}v,
\end{array}
\end{equation}
which was solved to obtain a perturbation growth rate for all possible perturbations $\delta=\delta_\text{re}+i\delta_\text{im}$. Vortex solitons are stable when $\delta_\text{re}=0$.
In Fig.~\ref{fig3} stable branches are shown in black, while unstable ones are shown in red. Vortex solitons with $m=1$ supported by $\mathcal{C}_4$ lattices 
\textcolor{black}{(according to the group theory arguments, this is the only possible charge of vortex soliton in the lattice with this discrete rotational symmetry)} are stable in the entire range of their existence, in both focusing and defocusing media, as shown in Fig. \ref{fig3}(a). In $\mathcal{C}_5$ lattice \textcolor{black}{(where only compact vortex states with topological charges up to $m=2$ are allowed on the disclination core)}, $m=1$ vortex solitons are stable [Fig. \ref{fig3}(b)], while $m=2$ ones are unstable [Fig.~\ref{fig3}(c)] in focusing medium ($g=+1$). Moreover, stability properties in $\mathcal{C}_5$ lattice change completely when the nonlinearity sign changes ($g=-1$), so that the families that were unstable in the focusing medium become stable in defocusing one, and vice versa \cite{kevrekidis2006high}. \textcolor{black}{It should be stressed that different stability properties of vortex solitons with lower and higher charges are typical for lattices with discrete rotational symmetry, but, at the same time, stability properties reported here are in clear contrast to those found in nontopological lattices \cite{kartashov2005soliton, law2009vort}, where, for example, in focusing medium only solitons with highest topological charges can be stable. We believe that this difference is connected with the fact that topological disclination vortex solitons appear in finite topological gap and on this reason their formation mechanism is different from that of conventional vortex solitons in semi-infinite gap.} Although vortex solitons in disclination lattices predominantly occupy the central sites within the disclination, they are fundamentally distinct from vortex solitons in nontopological ring-like waveguide arrangements, i.e.,  the structures obtained by excluding all waveguides that are not part of the disclination core~\cite{desyatnikov2011all}, as their stability characteristics are completely different from each other~\cite{suppmat}.

Figures~\ref{fig4}(a) and~\ref{fig4}(b) show examples of long-range stable propagation of $m=1$ vortex solitons in $\mathcal{C}_4$ and $\mathcal{C}_5$ disclination lattices, with dependencies of peak amplitude $A_\textrm{max}=\textrm{max}|\Psi|$ on distance $z$ and snapshots at different distances, obtained by direct numerical integration of Eq.~(1) using split-step fast Fourier method. Field modulus distributions in such states remain undistorted even after $z\sim 10^4$. At the same time, unstable $m=1$ states in defocusing medium [Fig.~\ref{fig4}(c)] and unstable $m=2$ state in focusing medium [Fig.~\ref{fig4}(d)] in $\mathcal{C}_5$ lattice show growing amplitude oscillations and typically lose vortical phase structure. The development of instability results in intensity oscillations of the spots, contributing significantly to the emission of radiation into the bulk of the lattice. This highlights the significance of the array surrounding the disclination core in defining the vortex soliton, distinct from the ring-like structures. Examples of stable and unstable evolution of vortices in $\mathcal{C}_{7,8}$ lattices are presented in~\cite{suppmat}.

\section{Conclusion and outlook}
We have shown that the symmetry of the disclination lattice plays a crucial role in the formation of vortex solitons of topological origin in the spectral topological gap. The formation of such states is facilitated by discrete rotational symmetry of the disclination lattice that simultaneously imposes strict restrictions on the topological charges of symmetric vortex solitons and determines their stability properties. In some cases, the symmetry can protect solitons from perturbations that would otherwise cause their decay, leading to unusual stability properties. Studying the interplay between nonlinearity and topology is essential for the development of new materials with desirable properties and the exploration of new opportunities for all-optical control of topological excitations, especially when they possess an orbital degree of freedom.


\begin{funding}
This work was supported by the Applied Basic Research Program of Shanxi Province (202303021211191), Y.V.K. acknowledges funding by the research project FFUU-2021-0003 of the Institute of Spectroscopy of the Russian Academy of Sciences and by the RSF grant 21-12-00096.
\end{funding}

\begin{authorcontributions}
All authors have accepted responsibility for the entire content of this manuscript and approved its submission.
\end{authorcontributions}

\begin{conflictofinterest}
Authors state no conflict of interest.
\end{conflictofinterest}

\begin{informedconsent}
Informed consent was obtained from all individuals included in this study.
\end{informedconsent}

\begin{ethicalapproval}
The conducted research is not related to either human or animals use.
\end{ethicalapproval}

\begin{dataavailabilitystatement}
The datasets generated during and/or analyzed during the current study are available from the
corresponding author on reasonable request.
\end{dataavailabilitystatement}



\end{document}


\title{Supplementary Material for \\ Vortex solitons in topological disclination lattices}

\author{Changming Huang}
\affiliation{Department of Physics, Changzhi University, Changzhi, Shanxi 046011, China}
\author{Ce Shang}
 \email{shang.ce@kaust.edu.sa}
\affiliation{King Abdullah University of Science and Technology (KAUST), Physical Science and Engineering Division (PSE), Thuwal 23955-6900, Saudi Arabia.}
\author{Yaroslav V. Kartashov}
\affiliation{Institute of Spectroscopy, Russian Academy of Sciences, 108840, Troitsk, Moscow, Russia}
\author{Fangwei Ye}
 \email{fangweiye@sjtu.edu.cn}
\affiliation{School of Physics and Astronomy, Shanghai Jiao Tong University, Shanghai 200240, China}
\date{\today}

\maketitle

In this Supplementary Material, we discuss the topological characterization of the disclination lattices and provide the results for linear spectra and vortex solitons in disclination lattices with higher, $\mathcal{C}_7$ and $\mathcal{C}_8$, discrete rotational symmetry. Such lattices are shown to support vortex solitons with topological charges up to $m=\pm 3$. The stability properties of such vortex solitons and the effect of the intrinsic loss of the waveguide array on their stability are discussed. Additionally, we conduct a systematic study comparing the stability of vortex solitons in a ring of waveguides with stability of vortex solitons supported by the disclination lattices, aiming to highlight the significance of the surrounding waveguide array in the stability of vortex solitons in disclination lattices.

\section*{1, Disclination lattices with $\mathcal{C}_7$ and $\mathcal{C}_8$ symmetry}
\begin{figure}[htbp]
\centering
\includegraphics[width=0.8\columnwidth]{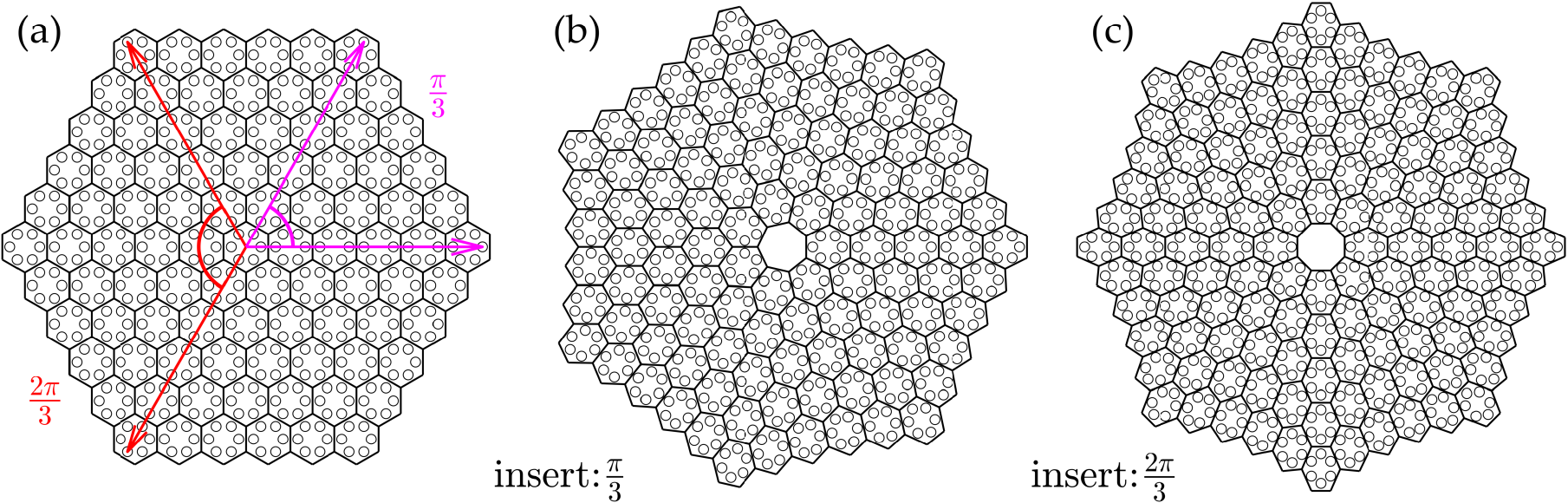}
\caption{Schematic illustration of the method of construction of the disclination lattices with higher discrete rotational symmetry. (a) Original hexagonal lattice structure. (b),(c) Disclination lattices with $\mathcal{C}_7$ (or $\mathcal{C}_8$) symmetry can be generated after insertion of a $\pi/3$ (or $2\pi/3$) sector into the hexagonal structure and subsequent compression of lattice cells.}
\label{figSI1}
\end{figure}

To create disclination lattices with discrete rotational symmetry higher than $\mathcal{C}_6$, we insert into the original hexagonal sample [Fig.~\ref{figSI1}(a)] the $n\pi/3$ sector, as shown in Fig.~\ref{figSI1}. We compress the cells accordingly so that the lattice accommodates the newly added sector. Specifically, when $n=1$, the resulting lattice exhibits $\mathcal{C}_7$ discrete rotational symmetry [Fig.~\ref{figSI1}(b)], while for $n=2$ the $\mathcal{C}_8$ structure can be obtained [Fig.~\ref{figSI1}(c)]. The declination core is clearly visible in the center of each such structure.

\section*{2, Linear modes in disclination lattices with $\mathcal{C}_7$ and $\mathcal{C}_8$ symmetry}
The increase of the order of discrete rotational symmetry $N$ leads to an increase in the number of linear modes localized at the disclination core of the $\mathcal{C}_N$ lattice. This phenomenon offers an opportunity for the generation of vortex states with higher topological charges. Therefore, it is essential to investigate the properties of linear localized disclination modes in $\mathcal{C}_7$ and $\mathcal{C}_8$ lattices.

\begin{figure*}[t]
	\centering
	\includegraphics[width=1\columnwidth]{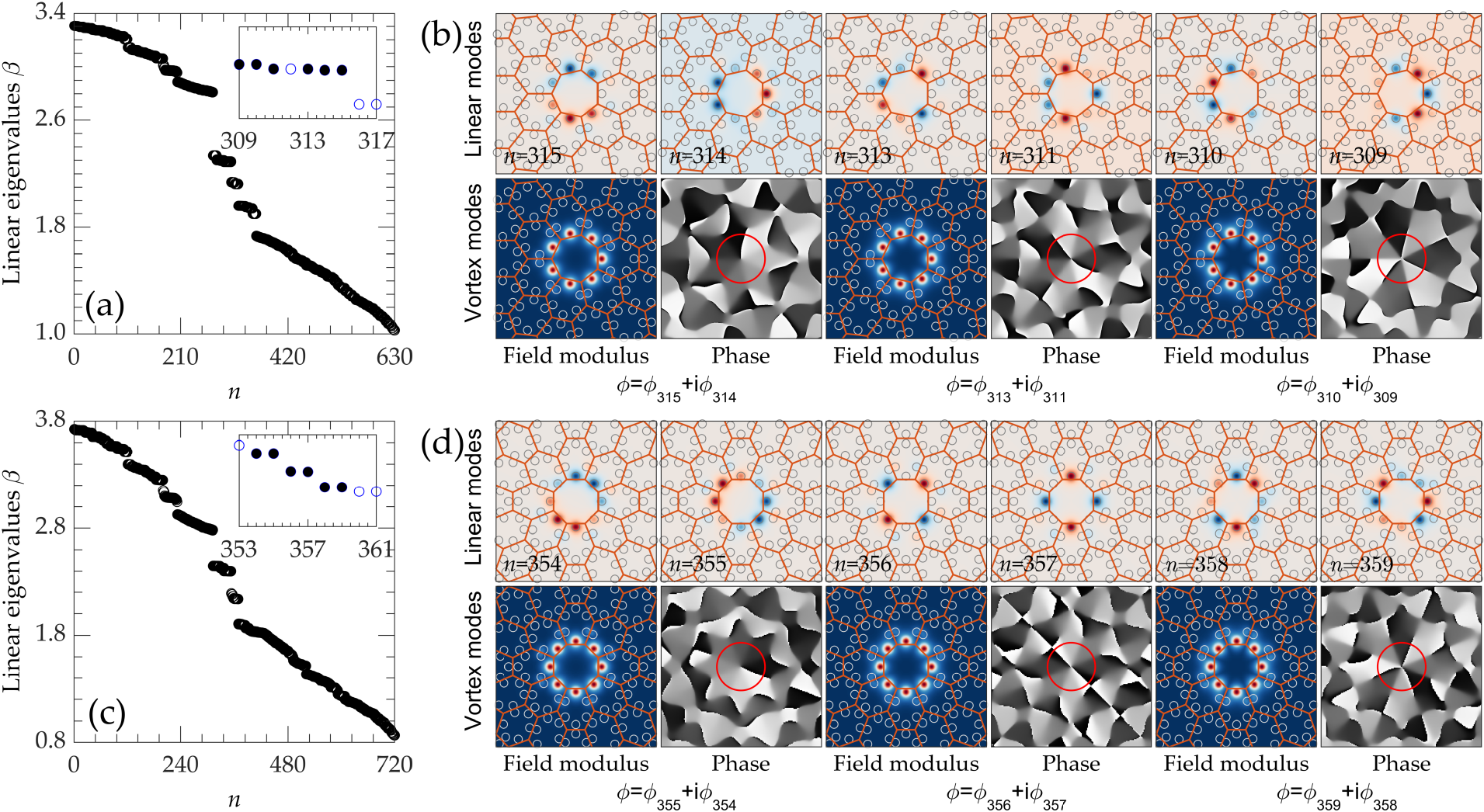}
	\caption{Linear eigenvalues, examples of linear disclination modes, and vortex modes that can be obtained in lattices with $\mathcal{C}_7$ and $\mathcal{C}_8$ symmetry. Linear eigenvalues of modes supported by the $\mathcal{C}_7$ (a) and $\mathcal{C}_8$ (c) lattice at $\gamma=1.72$ are shown. Solid dots in the insets indicate pairs of degenerate disclination modes that can be used for the construction of vortex states. The profiles of three pairs of degenerate eigenmodes supported by the $\mathcal{C}_7$ (b) or $\mathcal{C}_8$ (d) symmetric lattices (top row) and the corresponding vortex modes they generate (bottom row). In $\mathcal{C}_7$ lattice the degenerate pairs of disclination states are $\phi_{n=315,314}$, $\phi_{n=313,311}$, and $\phi_{n=310,309}$. In $\mathcal{C}_8$ lattice the degenerate pairs of disclination states are $\phi_{n=354,355}$, $\phi_{n=356,357}$, and $\phi_{n=358,359}$. The organge lines in (b) and (d) depict unit cells, while circles indicate the positions of the waveguides.}
	\label{figSI2}
\end{figure*}

These results once again confirm the conclusion drawn in the main text that the available charge of disclination vortex in $\mathcal{C}_N$ lattice is given by $m<N/2$ (for even $N$) and $m<(N+1)/2$ (for odd $N$). It is interesting to note that degenerate linear modes of disclination lattices that one can use for the construction of vortex states typically feature opposite symmetries with respect to some axis in the $(x,y)$ plane [for instance, $\phi_{n=315}$ and $\phi_{n=314}$ modes from Fig. \ref{figSI2}(b) are anti-symmetric (symmetric) with respect to $y=0$ axis].

\begin{figure*}[htbp]
	\centering
	\includegraphics[width=1\columnwidth]{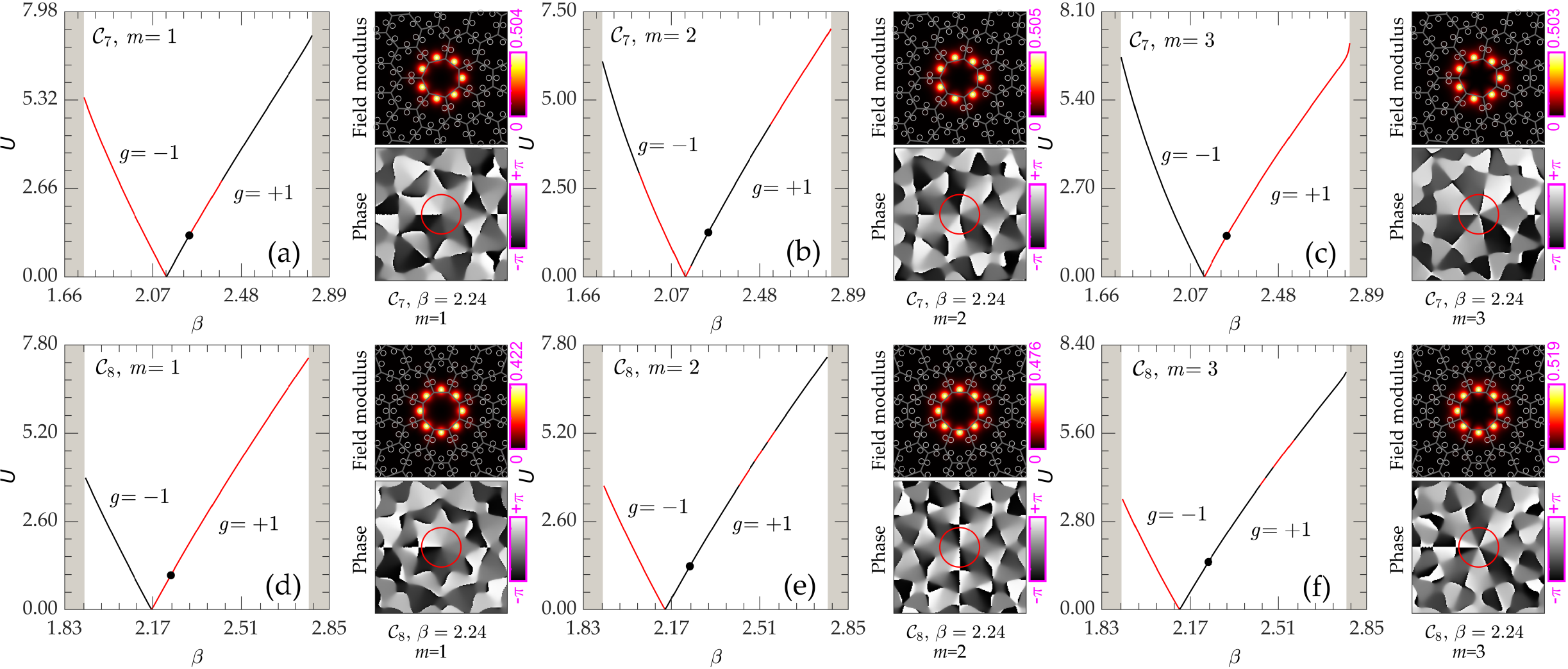}
	\caption{Vortex solitons supported by $\mathcal{C}_7$ and $\mathcal{C}_8$ symmetric lattices. The $U(\beta)$ curves for vortex gap solitons with $m=1$, $\mathcal{C}_7$ lattice (a), $m=2$, $\mathcal{C}_7$ lattice (b), $m=3$, $\mathcal{C}_7$ lattice (c), $m=1$, $\mathcal{C}_8$ lattice (d), $m=2$, $\mathcal{C}_8$ lattice (e), $m=3$, $\mathcal{C}_8$ lattice (f), are plotted. The gray regions in the $U(\beta)$ plots illustrate bulk bands. Stable families are indicated by black lines, while unstable families are indicated by red lines. Examples of vortex solitons are attached to the side of their respective $U(\beta)$ panels. The gray lines in the field modulus distributions depict the lattice unit cells, and the circles correspond to the position of the waveguides. Phase singularity in the center of each pattern is highlighted by the red circle.
   }
	\label{figSI3}
\end{figure*}
\section*{3, Vortex solitons in disclination lattices with $\mathcal{C}_7$ and $\mathcal{C}_8$ symmetry}
Vortex solitons with different topological charges bifurcating from linear topological vortex states have been found in both $\mathcal{C}_7$ and $\mathcal{C}_8$ lattices (see Fig. \ref{figSI3}). The $U(\beta)$ dependencies for these vortex solitons are qualitatively similar to those in $\mathcal{C}_4$ and $\mathcal{C}_5$ lattices. One can observe that nonlinear vortex states bifurcate from linear ones in both focusing and defocusing medium, with power gradually increasing toward the edge of the gap. The 7-peak and 8-peak vortex solitons are well localized around the disclination core, their vortical phase distributions with phase singularity in the center are highlighted by the red circles in phase structures shown in Fig. \ref{figSI3}. 

\begin{figure*}[htbp]
	\centering
	\includegraphics[width=1\columnwidth]{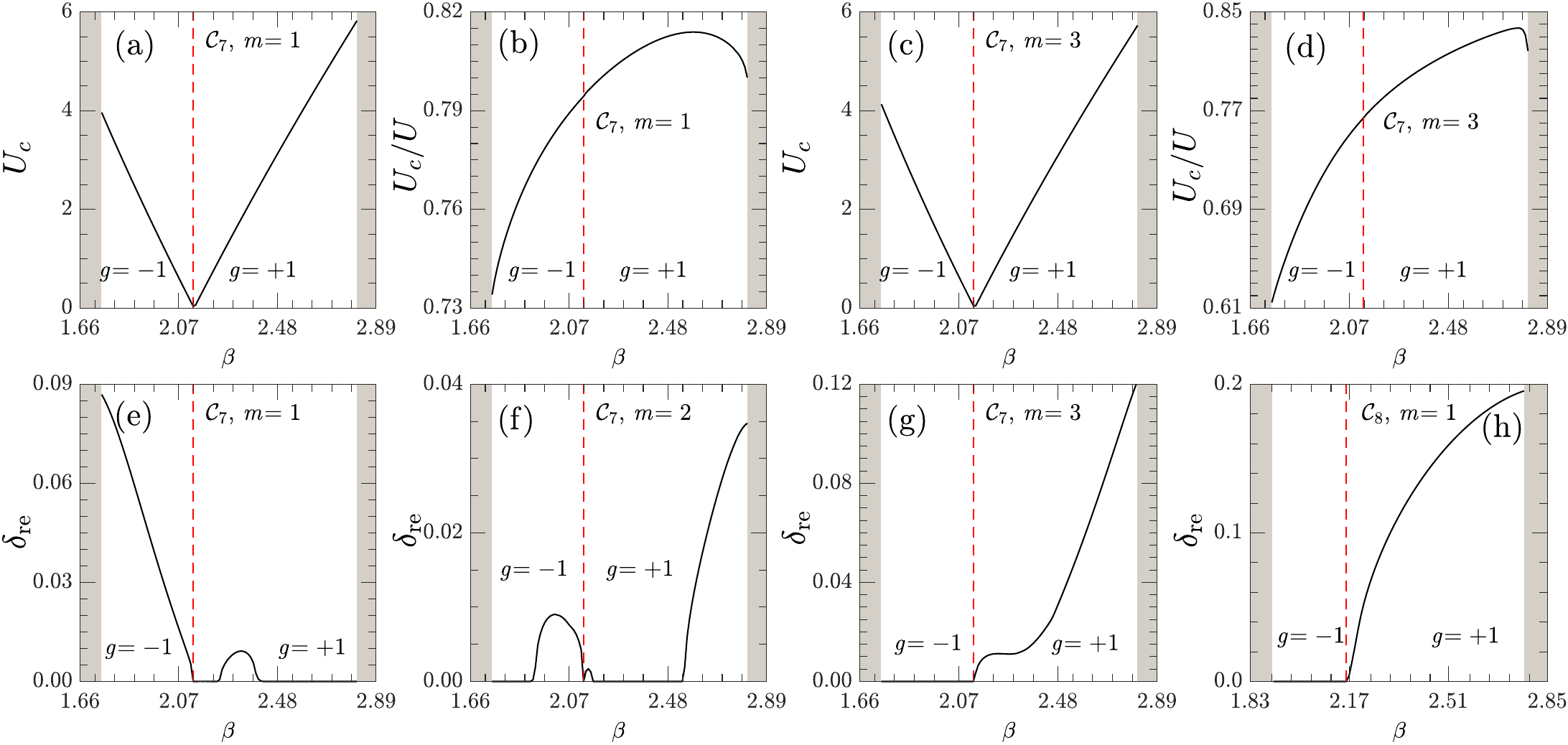}
	\caption{The dependence of  light power concentrated around the disclination core $U_c$ (or $U_c/U$)(the first line),  and the dependence of real part of the perturbation growth rate $\delta_\mathrm{re}$ (the second line), on the propagation constant $\beta$. The $U_c(\beta)$ and $U_c/U(\beta)$ curves for vortex solitons with $m=1$, $\mathcal{C}_7$ lattice (a,b), $m=3$, $\mathcal{C}_7$ lattice (c,d), are plotted. The $\delta_\mathrm{re}(\beta)$ curves for vortex solitons with $m=1$, $\mathcal{C}_7$ lattice (e), $m=2$, $\mathcal{C}_7$ lattice (f), $m=3$, $\mathcal{C}_7$ lattice (g), $m=1$, $\mathcal{C}_8$ lattice (h), are plotted. The red dashed line represents the boundary between focusing and defocusing nonlinearities.
 }
	\label{figSI4}
\end{figure*}

We present a quantitative analysis of the light power confinement around the disclination core, denoted as $U_c$ and defined as $U_c=\int_{0}^{2\pi}\int_{r_0-w}^{r_0+w}|\phi|^2rdrd\theta$. Here, $r_0$ represents the radius measured from the disclination core to the center of the central waveguides of the disclination lattices, $w$ is the waveguide width, and $\theta$ represents the angle in polar coordinates. The light power confinement $U_c$, and the ratio of the light power confinement around the disclination core to total power of the vortex soliton  ($U_c/U$) are depicted in  Figs.~\ref{figSI4}(a), \ref{figSI4}(b), \ref{figSI4}(c), and \ref{figSI4}(d). We observe that the light power confinement around the disclination core, $U_c$, exhibits the same trend of variation with $\beta$ as the $U(\beta)$ curve, while the fraction of power localized on disclination core $U_c/U$,  typically demonstrates an opposite variation with the strength of the nonlinearity in focusing nonlinearity compared to in defocusing nonlinearity. However, throughout their entire existence region, the fraction of power localized on disclination core remains above 60\%, indicating their nature as nonlinear states around the core of the disclination lattices.

\begin{figure*}[htbp]
	\centering
	\includegraphics[width=1\columnwidth]{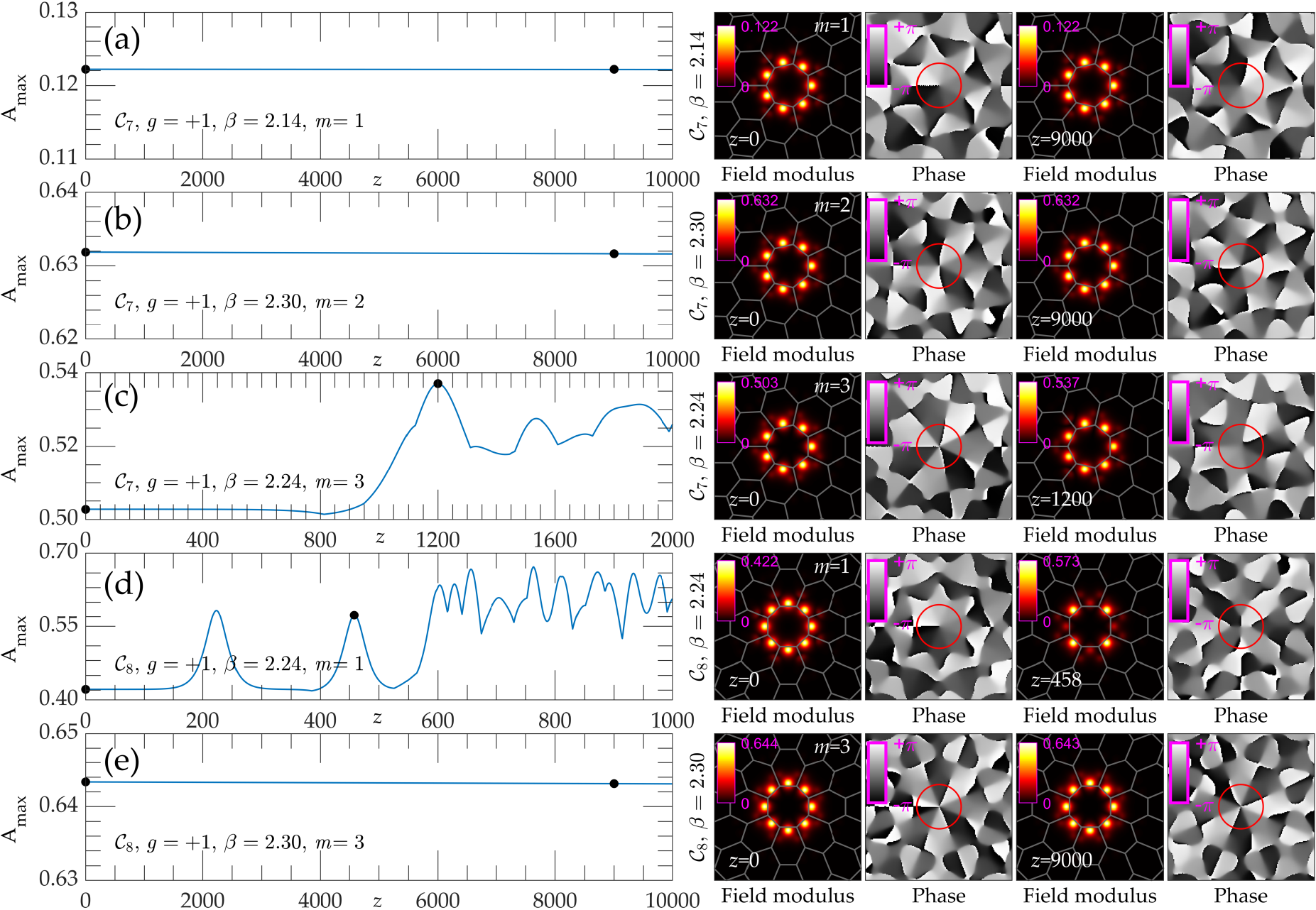}
	\caption{Propagation dynamics of vortex gap solitons in $\mathcal{C}_7$ and $\mathcal{C}_8$ symmetric lattices. The maximum amplitude $A_\text{max}$ of the field versus the propagation distance $z$ is shown on the left plots, while snapshots with field modulus and phase distributions corresponding to the black dots in the $A_\text{max}(z)$ plots are presented on the right. The evolution of stable vortex solitons is shown in panels (a), (b), and (c), while the unstable evolution is shown in panels (c) and (d). The gray lines in the field modulus panel depict the lattice cells, and the circles correspond to the position of the waveguides.  $m=1$, $\mathcal{C}_7$ lattice, $\beta=2.14$ in (a), $m=2$, $\mathcal{C}_7$ lattice, $\beta=2.3$ in (b), $m=3$, $\mathcal{C}_7$ lattice, $\beta=2.24$ in (c), $m=1$, $\mathcal{C}_8$ lattice, $\beta=2.24$ in (d), and $m=3$, $\mathcal{C}_8$ lattice, $\beta=2.3$ in (e).
}
	\label{figSI5}
\end{figure*}

\begin{figure*}[htbp]
	\centering
	\includegraphics[width=1\columnwidth]{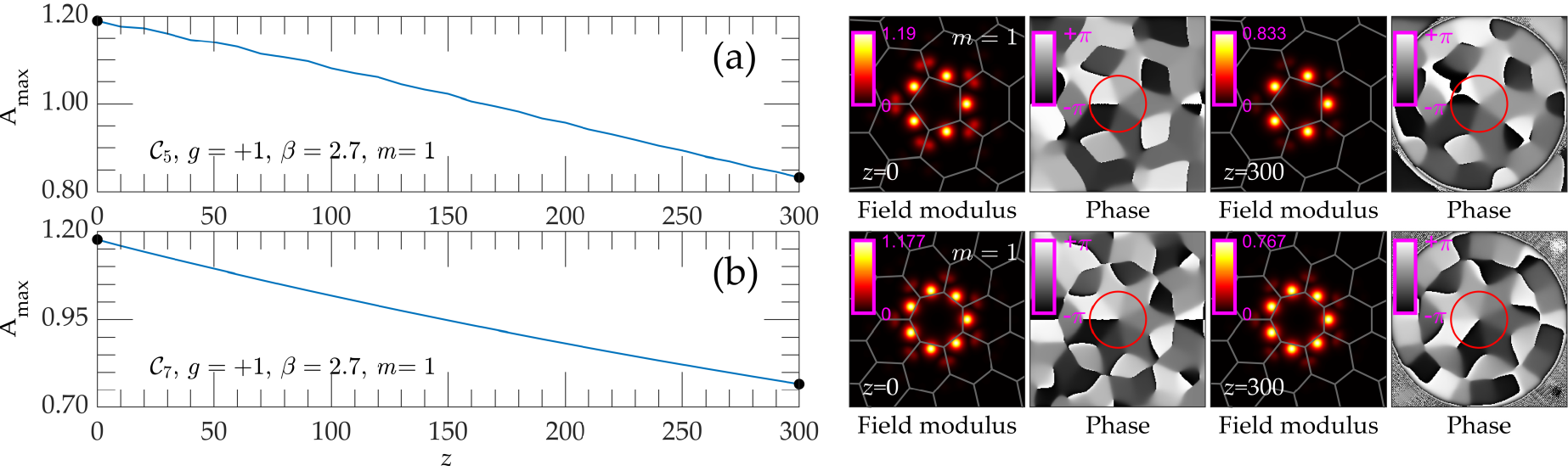}
	\caption{Examples of the propagation dynamics of vortex solitons with an absorption coefficient $\alpha=0.00131$. The maximum amplitude $A_\text{max}$ of the field versus the propagation distance $z$ is shown on the left plots, while snapshots with field modulus and phase distributions corresponding to the black dots in the $A_\text{max}(z)$ plots are presented on the right. The gray lines in the field modulus panel depict the lattice cells, and the circles correspond to the position of the waveguides. $m=1$, $\mathcal{C}_5$ lattice, $\beta=2.7$ in (a), $m=1$, $\mathcal{C}_7$ lattice, $\beta=2.7$ in (b).
    }
	\label{figSI6}
\end{figure*}

The stability picture for solitons in $\mathcal{C}_7$ and $\mathcal{C}_8$ lattices is more complex, which we determined through rigorous linear stability analysis. In Fig.~\ref{figSI3} stable and unstable branches are shown with black and red colors, respectively, in corresponding $U(\beta)$ curves. It is interesting that stable $m=1$ vortex solitons have been found in $\mathcal{C}_7$ lattice in focusing medium, while in $\mathcal{C}_8$ lattice such solitons are already unstable for $g=+1$. As before, stability properties for states with propagation constants not too far from the propagation constant of linear mode change when the sign of nonlinearity $g$ changes. Vortex solitons with $m=2$ and $m=3$ charges may have complex stability domains in $\mathcal{C}_8$ structure and for $g=+1$.

The examples of stable propagation of vortex solitons with different topological charges, which maintain their field modulus and phase distributions even after a considerable propagation distance ($z\sim10000$) in $\mathcal{C}_7$ and $\mathcal{C}_8$ disclination lattices are shown in Figs. \ref{figSI5}(a),~\ref{figSI5}(b), and~\ref{figSI5}(e). Unstable vortex solitons exhibit two distinct types of behavior: (1) the field modulus pattern maintains its shape but with small amplitude fluctuations, while the phase structure is lost after propagation over a certain distance [see Fig.~\ref{figSI5}(c)]; (2) bright spots in vortex profile show strong oscillations and after some propagation distance the field modulus distribution changes dramatically (the number of spots decreases) and vortical phase structure is lost as well [see Fig.~\ref{figSI5}(d)].

To further demonstrate that the vortex solitons evolve stably over experimentally feasible distances, we take into account the loss of waveguides. Such losses $\sim 0.1\mathrm{dB/cm}$ can be taken into account by including the term $-i\alpha\Psi$ into the right-hand side of Eq. (1), where the absorption coefficient is as small as $\alpha\approx0.00131$. The corresponding results are shown in Fig.~\ref{figSI6}. One can observe that the amplitude of the input vortex soliton gradually decreases with the propagation distance. However, within the range of an experimentally feasible sample length ($z\approx 175 \leftrightarrow 20~\mathrm{cm}$), the field modulus distribution changes only slightly (mainly due to decrease of peak amplitude of soliton) and representative phase structure is clearly conserved.

\begin{figure*}[b]
\centering
\includegraphics[width=0.85\columnwidth]{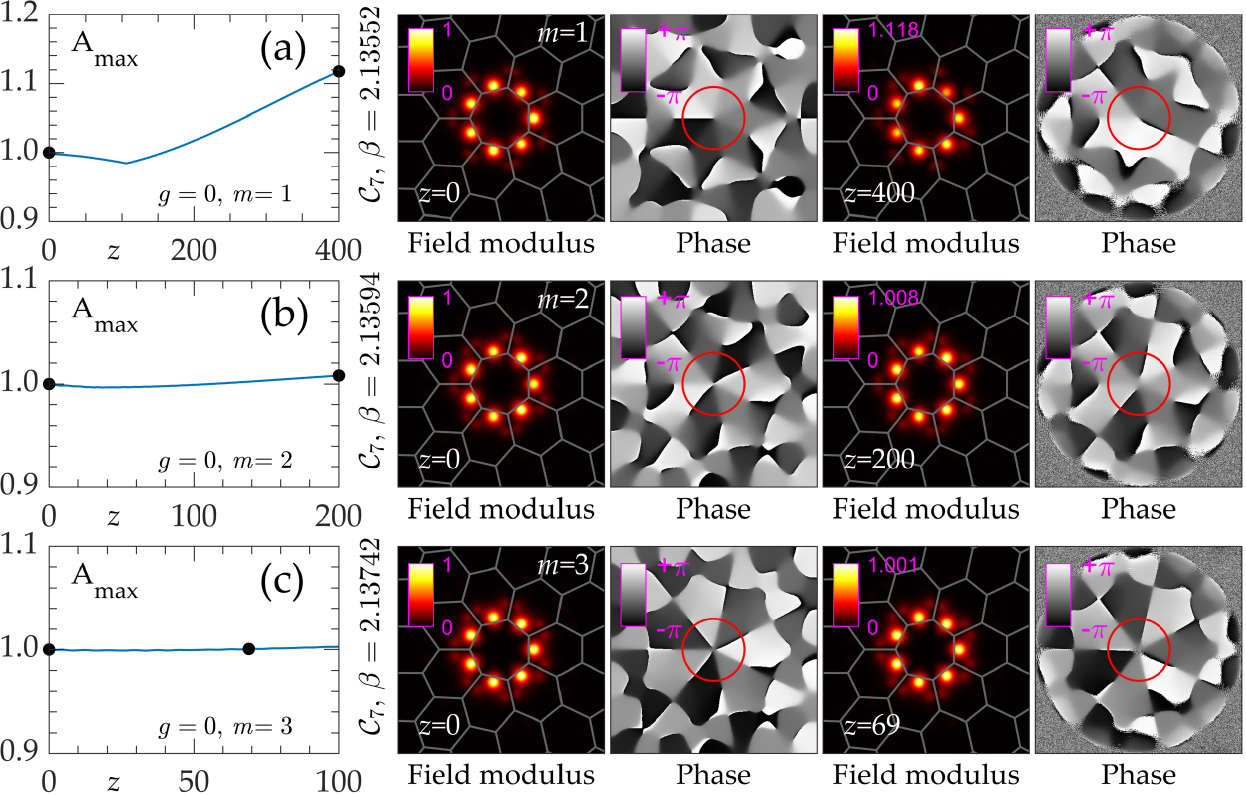}
\caption{\textcolor{black}{Propagation dynamics of linear vortex modes with $m=1, 2, 3$ in $5\%$ level noise  perturbed $\mathcal{C}_7$ symmetric lattices. The maximum amplitude $A_\text{max}$ of the field versus the propagation distance $z$ is shown on the left panels, while the snapshots with field modulus and phase distributions corresponding to the black dots in the $A_\text{max}(z)$ plots are presented on the right panels. The gray lines in the field modulus panel depict the lattice cells, and the circles correspond to the position of the waveguides.  
$m=1$, $\beta=2.13552$ in (a), $m=2$, $\beta=2.13594$ in (b), and $m=3$, $\beta=2.13742$ in (c). }
}
\label{figSI7}
\end{figure*}


\textcolor{black}{
It is interesting and relevant to consider the effect of perturbations that break the exact $\mathcal{C}_N$ discrete rotational symmetry of a structure on the evolution of vortex modes. To analyze the effects of these perturbations, we introduced disorder into all waveguides of the structure, including those located at the disclination core, by allowing their depths to change randomly within the interval $[p-\delta, p+\delta]$, where the disorder level $\delta$ is much smaller than $p$. We then propagated the linear vortex modes obtained in the unperturbed lattice with $\mathcal{C}_N$ symmetry in the lattice with the disorder, which does not possess any discrete rotational symmetry. The results are shown in Fig.~ S7, with the disorder level $\delta=0.05p$.  In all cases, we observed that the vortex modes with $m=1$ persist throughout the evolution [Fig. S7 (a)]. However, vortices with higher charges may exhibit splitting of the singularity at the center. This is clearly seen in Fig. S7 (b) and (c), where the vortex with $m=2$ quickly splits its double-folded singularity into two single singularities, and the vortex with $m=3$ splits its triple-folded singularity into three single singularities. Nevertheless, the removal of this degeneracy in phase singularity does not result in the destruction of the vortex state, as evidently shown in all the propagation simulations of Fig.~S7.}


\begin{figure*}[b]
	\centering
	\includegraphics[width=0.95\columnwidth]{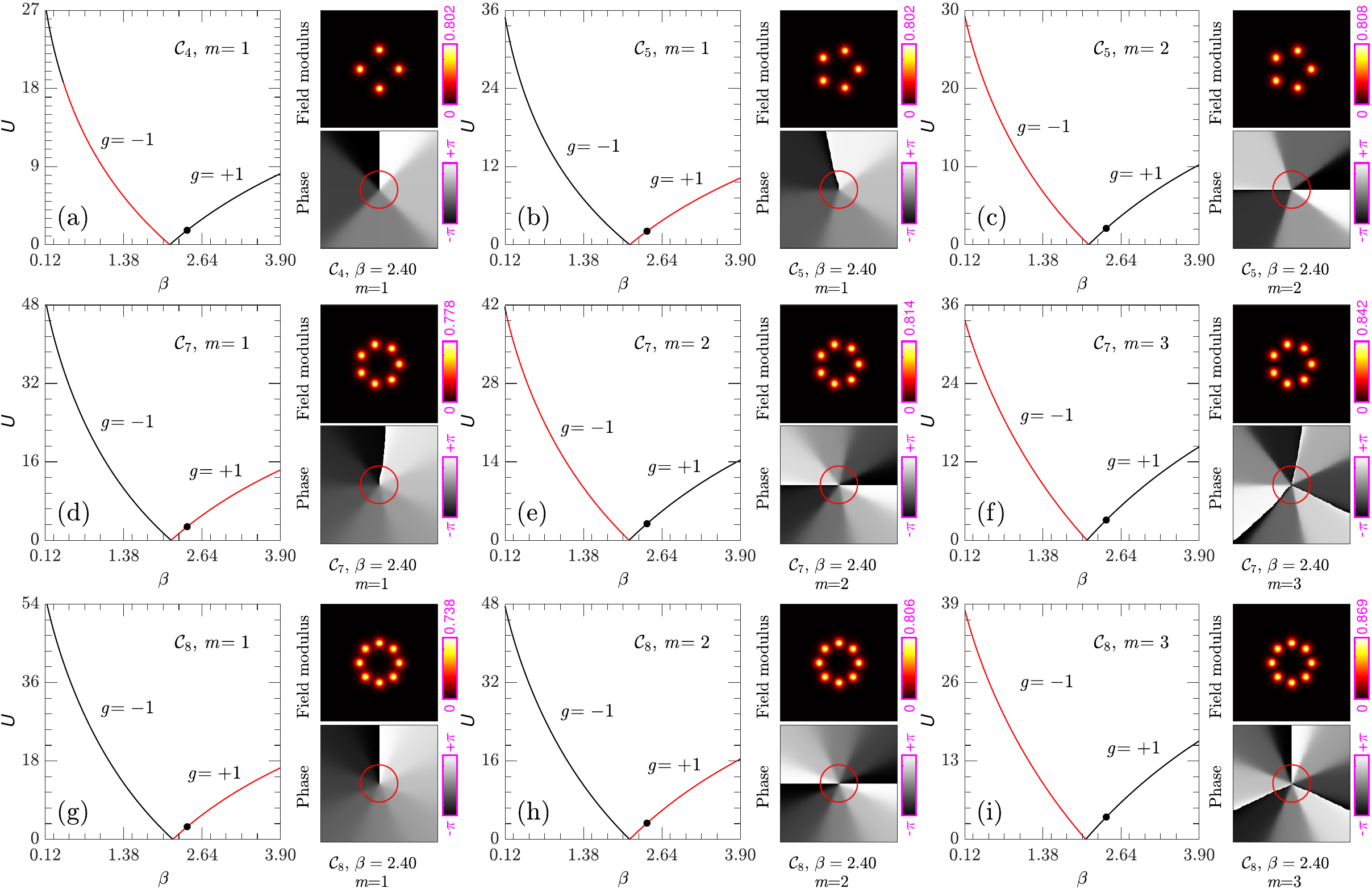}
	\caption{Vortex solitons supported by ring-like lattices.
 The $U(\beta)$ curves for vortex solitons with $m=1$, $\mathcal{C}_4$ lattice (a), $m=1$, $\mathcal{C}_5$ lattice (b), $m=2$, $\mathcal{C}_5$ lattice (c), $m=1$, $\mathcal{C}_7$ lattice (d), $m=2$, $\mathcal{C}_7$ lattice (e), $m=3$, $\mathcal{C}_7$ lattice (f), $m=1$, $\mathcal{C}_8$ lattice (g), $m=2$, $\mathcal{C}_8$ lattice (h), $m=3$, $\mathcal{C}_8$ lattice (i), are plotted.
 Stable families are indicated by black lines, while unstable families are indicated by red lines. Examples of vortex solitons are attached to the side of their respective $U(\beta)$ panels. Phase singularity in the center of each pattern is highlighted by the red circle.}
	\label{figSI8}
\end{figure*}
\begin{figure*}[t]
	\centering
	\includegraphics[width=0.75\columnwidth]{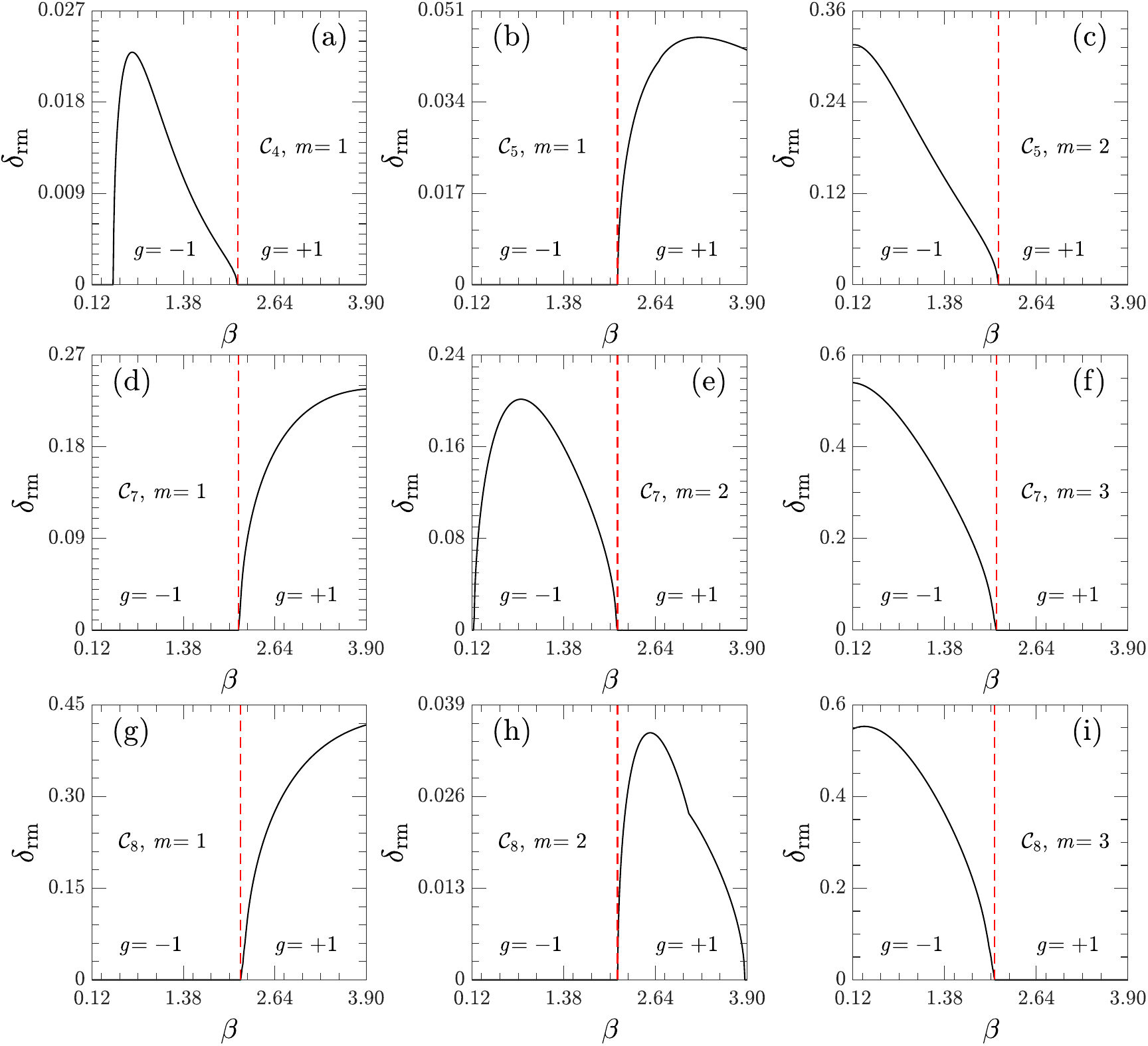}
	\caption{The dependence of real part of the perturbation growth rate $\delta_\mathrm{re}$ on the propagation constant $\beta$ for vortex solitons supported by ring-like lattices. The $\delta_\mathrm{re}(\beta)$ curves for vortex solitons with $m=1$, $\mathcal{C}_4$ lattice (a), $m=1$, $\mathcal{C}_5$ lattice (b), $m=2$, $\mathcal{C}_5$ lattice (c), $m=1$, $\mathcal{C}_7$ lattice (d), $m=2$, $\mathcal{C}_7$ lattice (e), $m=3$, $\mathcal{C}_7$ lattice (f), $m=1$, $\mathcal{C}_8$ lattice (g), $m=2$, $\mathcal{C}_8$ lattice (h), $m=3$, $\mathcal{C}_8$ lattice (i), are plotted. The red dashed line represents the boundary between focusing and defocusing nonlinearities.
 }
	\label{figSI9}
\end{figure*}

\section*{4, Vortex solitons in ring-like lattices}
In this section, we conduct a systematic study on the stability of vortex solitons in a ring of waveguides and compare it to that of vortex solitons in disclination lattices. To create the ring-like waveguide, we remove all waveguides from the disclination lattices that are not part of the disclination core, as shown in Fig.~\ref{figSI8}. Our study considers both focusing and defocusing nonlinearities.

We observe two key differences between vortex solitons in ring-like waveguides and those in disclination lattices. In the ring-like waveguides, as shown in Fig.~\ref{figSI8}, the domain of existence for vortex solitons is much broader compared to that of gap vortex solitons in topological disclination lattices. This is because the ring-like structure lacks band structures that could determine the existence domain of the vortex solitons. Additionally, the power of these vortex solitons is entirely confined within the ring-like waveguide structure. As a result, the vortex solitons in the ring-like waveguides always have monotonically decaying tails, which is distinct from their counterparts in disclination lattices where the gap vortices feature oscillating tails that penetrate into the lattice bulk.

To ensure a fair comparison between the ring structure and the disclination lattice, we set the ring-like lattice to have the same parameters as the disclination lattice (i.e., $p=8$ and $w=0.5$). Our linear stability analysis reveals that vortices with $m=1$ (or $m=2$) in $\mathcal{C}_4$ (or $\mathcal{C}_5$) symmetric ring-like lattices are stable throughout their entire existence domain under the condition of focusing nonlinearity [see Figs. \ref{figSI9}(a) and~\ref{figSI9}(c)]. Moreover, $\mathcal{C}_7$ symmetric ring-like lattices support stable vortices with $m=1$ and $m=2$ [see Figs. \ref{figSI9}(e) and~\ref{figSI9}(f)], while $\mathcal{C}_8$ symmetric ring-like lattices support stable vortices with $m=3$ [see Fig. \ref{figSI9}(i)].
In the focusing medium, stable vortex solitons with $m=1$ have not been found in $\mathcal{C}_{N}$ ($N>4$) symmetric ring-like lattices [see Figs. \ref{figSI9}(b),~\ref{figSI9}(d),~and \ref{figSI9}(g)]. Similar conclusions have also been presented in Refs.~\cite{kartashov2005soliton,desyatnikov2011all}. In the ring-like lattices, the flip of vortex stability under a change of sign of nonlinearity has been observed \cite{kevrekidis2006high}. Comparing the stability of the vortex solitons in such ring-like structures with their counterpart in disclination latices, one immediately finds the stability between these two are entirely different, in other words, the stability of the vortex solitons in disclination lattices have been fundamentally altered by the presence of the waveguide arrays surrounding the core of the disclination lattices.


\section*{5, Topological index}
 The crystalline topology can be deduced from the symmetry indicators  (band representations) \cite{Bradlyn2017}. For a higher-order topological insulator with a hexagonal unit cell, the primary topological index is \cite{PhysRevB.99.245151} 
\begin{equation}
\chi^{(6)}=\left(\left[M_1^{(2)}\right],\left[K_1^{(3)}\right]\right).
\end{equation}
where $\left[\Pi_{p}^{(n)}\right] \equiv \# \Pi_{p}^{(n)}-\# \Gamma_{p}^{(n)}$.  Here $\# \Pi_{p}^{(n)}$ is the number of occupied bands at the high-symmetry point $\Pi\ (=\mathrm{M},\mathrm{K})$ with the $C_n$ rotation eigenvalues $e^{2\pi \mathrm{i}(p-1)/n}\ (p = 1, ..., n)$. A disclination is characterized by the net translation (denoted by the Burgers vector $B$) and net rotation (denoted by the Frank angle $\Omega$) accumulated under parallel transport of a vector along a loop enclosing the core. For the topologically nontrivial case ($\gamma=d_1/d_2>1$), one can find $\chi^{(6)}=(2,0)$. While for the topologically trivial case ($\gamma=d_1/d_2<1$), $\chi^{(6)}=(0,0)$. The secondary topological index of the disclination is given by 
\begin{equation}\label{dis}
Q=\frac{\Omega}{2 \pi}(\frac{3}{2}\left[M_1^{(2)}\right]-\left[K_1^{(3)}\right])\bmod 1,
\end{equation}
yielding $Q= 0$ for a trivial phase and $Q ={3\Omega}/{2 \pi} \bmod 1$ for a nontrivial phase.

%